\documentclass[aps,prc,twocolumn,showpacs,floatfix,nofootinbib]{revtex4-1}
\usepackage{amsmath,graphicx,color}
\usepackage[colorlinks,linkcolor=blue,citecolor=blue]{hyperref}

\def\E{{\mathcal E}}

\def\Eq#1{Eq.~(\ref{#1})}

\def\Fig#1{Fig.~\ref{#1}}
\def\Sect#1{Section~\ref{#1}}

\def\bra{\langle}
\def\ket{\rangle}

\def\be{\begin{equation}}
\def\ee{\end{equation}}
\def\bea{\begin{eqnarray}}
\def\eea{\end{eqnarray}}

\begin{document}

\title{Response relation in Pb-Pb and p-Pb collisions at 5.02 TeV}
\author{De-Xian Wei}
\email{dexianwei@gxust.edu.cn}
\author{Li-Juan Zhou}
\author{Xin-Fei Li}
\affiliation{School of Science, Guangxi University of Science and Technology, Liuzhou, 545006, China}

\begin{abstract}
We carry out simulations using a multi-phase transport (AMPT) model to describe the response relation between $v_2$ and $\varepsilon_2$
in Pb-Pb and p-Pb collisions at $\sqrt{s_{NN}}=5.02$ TeV, respectively.
To simulate such relation, two methods have been introduced in the calculation: one is the directed response (DR) method, which correlates the outgoing particles with the initial anisotropy directly, and the other one is the cumulants response (CR) method, which is constructed from a cumulants correlation between outgoing particles.
Based on calculations of the DR and CR methods, the response relations as a function of the transverse momentum are both shown in Pb-Pb and p-Pb collisions.
By comparing the DR and CR methods, we found that the linear response relations are almost identical in all the present collisions. Similar results of linear+cubic response relations
are also shown in the higher multiplicity systems, and it has become a significant difference in the lower multiplicity systems, i.e., the peripheral Pb-Pb collisions and p-Pb collisions.
Throughout the whole $p_{T}$-dependent simulations, the $\kappa_2$ in the linear response and in the linear+cubic response are almost identical,
except for in the lower multiplicity systems by the DR method.
If one implements a pseudorapidity gap by the CR calculation, the $p_{T}$-dependent and $\eta$-independent response relations are similarly shown in peripheral Pb-Pb systems and p-Pb systems, which may imply that a collective response exists in the most central p-Pb collisions.
These collective behaviors are dominantly produced on the stage of the medium expansions.
\end{abstract}
\maketitle

\section{Introduction}
One remarkable achievement in high-energy heavy-ion experiments is the creation of a fluidlike quark-gluon system: the quark-gluon plasma (QGP).
The collective flow behavior plays an important role in probing the fluidlike QGP and is studied in the long-range azimuthal correlations of particles emitted in
relativistic nucleus-nucleus collisions at the Brookhaven National Laboratory Relativisitic Heavy Ion Collider (BNL RHIC)~\cite{Abelev:2009lrr} and at the European Organization for Nuclear Research Large Hadron Collider (CERN LHC)~\cite{ATLAS:2012mot}.
These collective flows can be described well by the hydrodynamic model~\cite{McDonald:2017hpf} and has shown that the collective flow behavior in particular is sensitive to the initial fluctuation states.
To study these fluctuations of initial states, various hydrodynamic and transport model have demonstrated an approximate response relation between the final harmonic flow $v_{n}$ and the initial eccentricity $\varepsilon_{n}$~\cite{Alver:2010tfi,Noronha-Hostler:2015dbi,Yan:2017ivm,De:2018hri}. These response relations are presented as dependent on transverse momentum $p_{T}$, pseudorapidity $\eta$ and system size, which shows it is crucial to understand the fluctuations of initial states~\cite{Ma:2016ipc,Hui:2019pdh}. Furthermore, it had led a nonlinear response for the $v_{n}$ with $\varepsilon_{n}$ by a hydrodynamic model~\cite{Yan:2017ivm}, and such nonlinear response is induced by the initial fluctuations. However, there is still a lack of a clear figure on the $p_{T}$-differential nonlinear response both in the hydrodynamic and transport models. \\

In recent years, similar long-range collective azimuthal correlations have also been observed in events with high final-state particle multiplicity
in proton-proton~\cite{CMS:2017efc}, proton-nucleus~\cite{CMS:2015efc,ATLAS:2018mol},
and lighter nucleus-nucleus collisions~\cite{PHENIX:2018mom},
where it naturally raises the question of whether a fluidlike QGP is created in these much smaller systems~\cite{Nagle:2018ssc}.
Even though the response relations have been studied systemically by hydrodynamic or transport model, but it is still incomplete in small systems,
due to the collective flow in such a small system size still being debated~\cite{Dusling:2016ncp}.
It has been argued that the system size is too small and the lifetime is too short for the matter
in a small system to hydrodynamize and approach local isotropization~\cite{Schenke:2017ooc}.
In this paper, we focus on the collective flow response to its initial fluctuating geometry states in p-Pb collisions.
This work follows our previous work, which has dealt with the response relations in Pb-Pb systems~\cite{De:2018hri}.
The main purpose of this paper is to address a particular picture of response theory in p-Pb systems:
to extract the collective response relations and to compare them with Pb-Pb systems.

This paper is organized as follows: In \Sect{sec:sec2} we briefly describe a multi-phase transport (AMPT) model~\cite{Lin:2004en} which used in
the present simulations. The numerical results of response relations for $v_2$
are presented in \Sect{sec:sec3}, where we emphasize the collective response and the residual noise in Pb-Pb and p-Pb collisions.
All the produced charge particles in the calculation are chosen with 0.3 $< p_{T} <$ 2.0 GeV and $|\eta|<$2.5.
Finally, we will summarize the main results in \Sect{sec:sum}. We use natural unit $k_B=c=\hbar=1$.

\section{A multi-phase transport (AMPT) model}
\label{sec:sec2}
The AMPT model is a hybrid transport model for high-energy heavy ion collisions~\cite{Lin:2004en}.
The AMPT model can be produced particles in difference stages, from initial to final states.
It not only includes the medium matter but also includes the nonflow effects which are produced in hard scattering and hadron decay.
To study the hydrodynamic property, we need a good physical quantity to describe the collective property of the medium, one is the harmonic flow.
If there exists a nonzero hydrodynamic response of harmonic flow, there implies that a fluid medium has been produced in the system.
To extract the hydrodynamic response, one must clearly eliminate the contribution of residual noise on the response relations.
To theoretically investigate these collective behaviors, AMPT is an appropriate tool for the response analysis.
More thorough discussions of the AMPT model can be found in Ref.~\cite{Lin:2004en}. \par

In this paper, the Lund
string fragmentation parameters, $a=0.5$, $b=0.9$ GeV$^{-2}$,
$\alpha_{s}$ =0.33 and $\mu$ = 3.2 fm$^{-1}$ are taken as in Ref.~\cite{De:2018hri}.
Throughout this paper, our results and analyses are mostly obtained based on
AMPT simulations with respect to Pb-Pb and p-Pb collisions at $\sqrt{s_{NN}}=5.02$ TeV.

\section{Response relation in AMPT}
\label{sec:sec3}

Staring from the harmonic flow $V_n$ estimators, which are studied in Ref.~\cite{Alver:2010gr},
\be
\label{eq:vn}
V_n = v_n e^{in\Psi_n} \equiv \int \frac{d\phi}{2\pi} e^{in\phi_p} f(\phi_p)\,.
\ee
where the magnitude $v_n$ and phase $\Psi_n$ fluctuate on an event-by-event basis.
$V_{n}$ is found to be very sensitive both to the event-by-event fluctuating initial eccentricity,
the transport properties, and their equation of state~\cite{Heinz:2013cfa,ALICE:2016cef,Nie:2019ioi}.
The initial eccentricity $\E_n$ is defined with respect to the initial-state energy density profile $\rho(\vec x_\perp, \tau_0)$
as~\cite{Teaney:2010vd}
\be
\label{eq:en}
\E_n=\varepsilon_n e^{in\Phi_n}\equiv -{\int d^2\vec x_\perp \rho(\vec x_\perp, \tau_0) r^n e^{in\phi}
\over
\int d^2\vec x_\perp r^n\rho(\vec x_\perp, \tau_0)}\,~~(n\geq2).
\ee
Since $\E_n$ is defined, harmonic flow $V_n$ can be expanded
with respect to initial eccentricity $\E_n$, and a series of response relations can be obtained.
In this work, we focus on the second-order harmonic for the response analyses.\\

Noronha-Hoster and his collaborators first studied the initial fluctuations driving the cubic response by a hydrodynamic model~\cite{Noronha-Hostler:2015dbi}, and similar results of AMPT are also shown in our previous work~\cite{De:2018hri}. The initial fluctuation-driven response relations between the $V_2$ and $\E_2$ is expressed as
\be
\label{eq:resp}
V_2=\kappa_2 \E_2 + \kappa_2' \varepsilon_2^2\E_2 + \delta_2.
\ee .
For a rotational symmetry condition, the leading-order term is a linear response
proportional to $\E_2$, with $\kappa_2$ the linear response coefficient determined by the medium dynamical
expansion.  The next leading-order contribution is a cubic term.
Both the linear and cubic terms are influenced by fluctuations where the centrality is larger than 30\% in Pb-Pb collisions, as noted by hydrodynamic simulations~\cite{Noronha-Hostler:2015dbi}.
Note that $V_2$ calculated from \Eq{eq:vn}
sums up all charged particles in the present $p_{T}$ and $\eta$ ranges in entire events by the event-plane method. As a consequence, collective flows and others (denoted residual noise, $\delta_{2}$) are included in the considered range. By minimizing the effect of additional residual noise $\delta_2$, one solves $\kappa_2$ and $\kappa_2'$,
here is denoted as the directed response (DR)~\cite{Noronha-Hostler:2015dbi,De:2018hri},
\begin{subequations}
\label{kappaprime}
\begin{align}
\kappa_2&=\frac{
{\rm Re}\left(
\langle \varepsilon_2^6\rangle\langle  V_2\E_2^*\rangle
-\langle \varepsilon_2^4\rangle\langle
V_2\E_2^*\varepsilon_2^2\rangle\right)}
{\langle \varepsilon_2^6\rangle\langle \varepsilon_2^2\rangle-\langle
 \varepsilon_2^4\rangle^2},\\
\kappa'_2&=\frac{
{\rm Re}\left(
-\langle \varepsilon_2^4\rangle\langle  V_2\E_2^*\rangle
+\langle \varepsilon_2^2\rangle\langle
V_2\E_2^*|\varepsilon_2|^2\rangle\right)}
{\langle \varepsilon_2^6\rangle\langle \varepsilon_2^2\rangle-\langle
  \varepsilon_2^4\rangle^2}.
\end{align}
\end{subequations}
where brackets $\bra \ldots\ket$ indicates the present particles are averaged over events.

If one only considers a linear response in \Eq{eq:resp},  $V_2=\kappa_2 \E_2$, it can get a solution as $\kappa_2=\frac{\langle  V_2\E_2^*\rangle}{\langle \varepsilon_2^2\rangle}$ (named $\kappa_2$ DR-linear).


To subtract nonflow effects in the event-averaged harmonic flow, pseudo-rapidity gap and multiparticle cumulants have been used for the calculations~\cite{Chatrchyan:2014kba,Acharya:2018zuq,Aad:2014vba}. Following this, a similarly subtracted residual term is introduced.
Based on the response relations in \Eq{eq:resp}, in which two particles correlate with pseudorapidity gap in Ref.~\cite{Chatrchyan:2013mot} and four-particle cumulants in Ref.~\cite{Yan:2014aad},
we find from the two-particle and four-particle correlations [denoted as cumulants response (CR)~\cite{De:2018hri}]
\begin{subequations}
\label{eq:nonflow_v2}
\begin{align}
v_{2}\{2,|\Delta\eta|\} &= \kappa_{2}\varepsilon_{2}\{2\} + \kappa_{2}^{'}\frac{\langle\varepsilon_{2}^{4}\rangle}{\langle\varepsilon_{2}^{2}\rangle}\varepsilon_{2}\{2\}\,, \\
v_{2}\{4\} &= \kappa_{2}\varepsilon_{2}\{4\}+ \kappa_{2}^{'}\frac{2\langle\varepsilon_{2}^{2}\rangle\langle\varepsilon_{2}^{4}\rangle-\langle\varepsilon_{2}^{6}\rangle}{2\langle\varepsilon_{2}^{2}\rangle^{2}-\langle\varepsilon_{2}^{4}\rangle}\varepsilon_{2}\{4\}\,.
\end{align}
\end{subequations}
Here, unlike in the practice of experimentally examining the multiparticle cumulant flow with only one particle in the present $p_{T}$ range, in the simulation of evaluating $v_{2}\{n\}$ ($n=$2,4), we placed all particles in the same $p_{T}$ range, e.g., two particles between the $\Delta\eta$ of $v_{2}\{2,|\Delta\eta|\}$ have the same $p_{T}$ value. In writing \Eq{eq:nonflow_v2}, we have assumed that a pseudo-rapidity gap is sufficient to
take out the residual noise contribution, i.e., $\delta_2$, in $v_2\{2\}$.
\Eq{eq:nonflow_v2} then allows us to solve $\kappa_2$
and $\kappa_2'$, without residual noise, as
\begin{subequations}
\label{eq:nonflow_solue}
\begin{align}
\kappa_{2} &= \frac{v_{2}\{2,|\Delta\eta|\} B-v_{2}\{4\} A}{\varepsilon_{2}\{2\} B-\varepsilon_{2}\{4\} A},  \\
\kappa_{2}^{'} &= \frac{-(v_{2}\{2,|\Delta\eta|\}\varepsilon_{2}\{4\}-v_{2}\{4\}\varepsilon_{2}\{4\})}{\varepsilon_{2}\{2\} B-\varepsilon_{2}\{4\} A}.
\end{align}
\end{subequations}
where
\begin{eqnarray}
\label{eq:nonflow_solue2}
A = \frac{\langle\varepsilon_{2}^{4}\rangle }{\langle\varepsilon_{2}^{2}\rangle} \varepsilon_{2}\{2\},
B = \frac{2\langle\varepsilon_{2}^{2}\rangle\langle\varepsilon_{2}^{4}\rangle-\langle\varepsilon_{2}^{6}\rangle}{2\langle\varepsilon_{2}^{2}\rangle^2-\langle\varepsilon_{2}^{4}\rangle} \varepsilon_{2}\{4\}. \nonumber
\end{eqnarray}

If one substitutes $v_{2}\{2,|\Delta\eta|\} = \kappa_{2}\varepsilon_{2}\{2\}$ in \Eq{eq:nonflow_v2}, then one can get the linear solution of the cumulants response, as $\kappa_{2}=v_{2}\{2,|\Delta\eta|\}/\varepsilon_{2}\{2\}$ (named $\kappa_2$ CR-linear).

\begin{figure*}
\begin{center}
\includegraphics[height=0.30\textwidth]{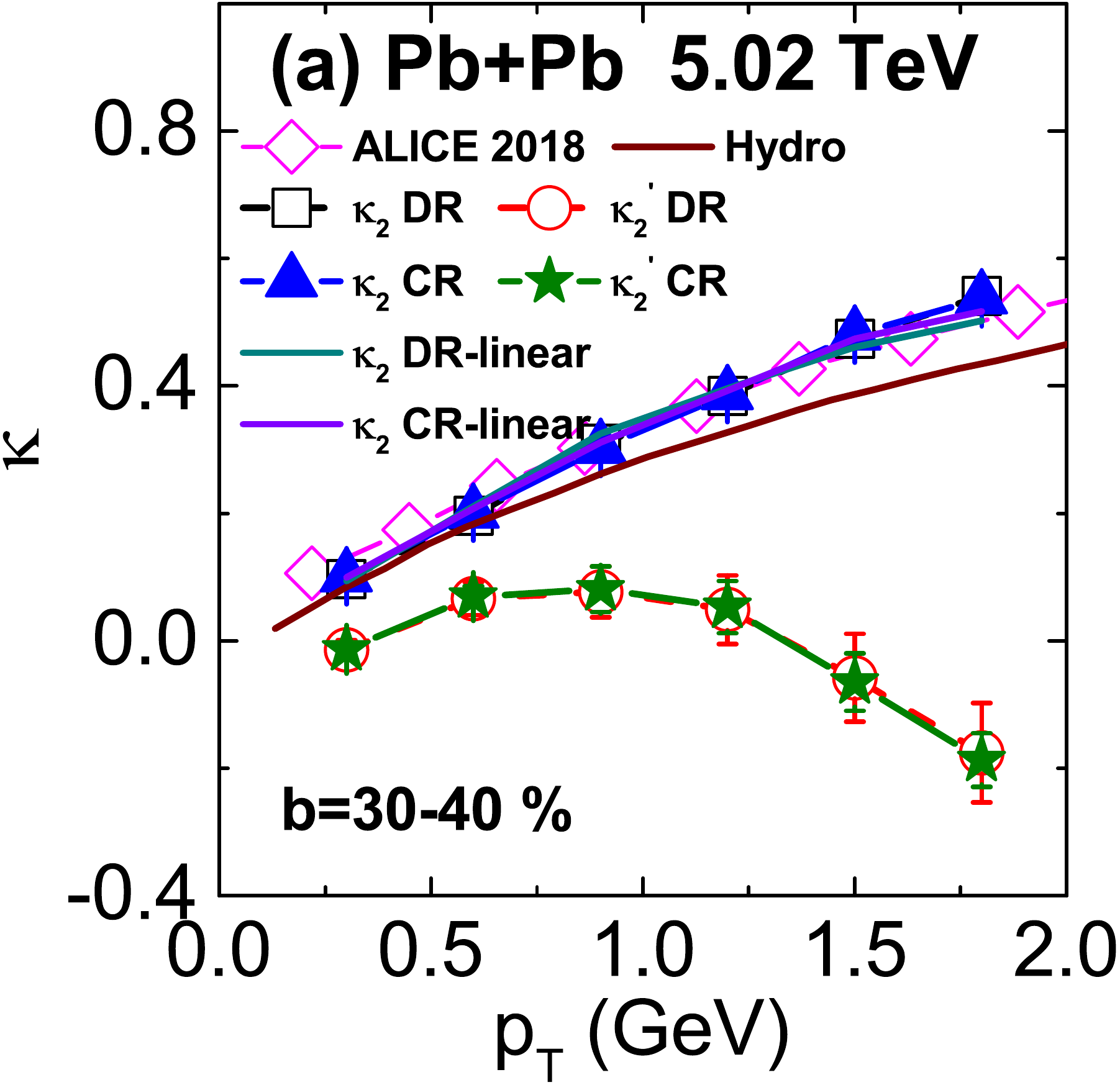}
\includegraphics[height=0.30\textwidth]{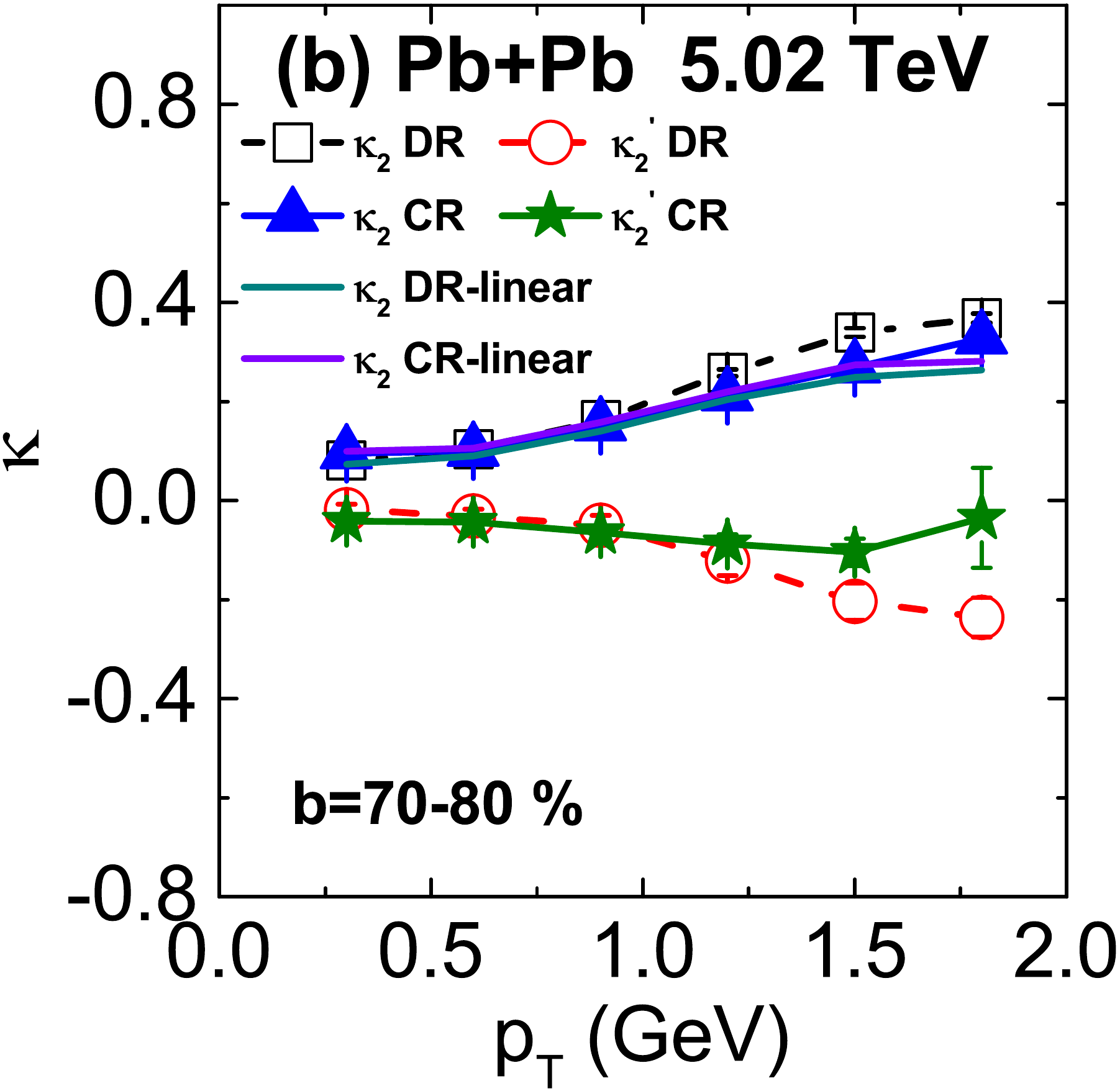}
\includegraphics[height=0.30\textwidth]{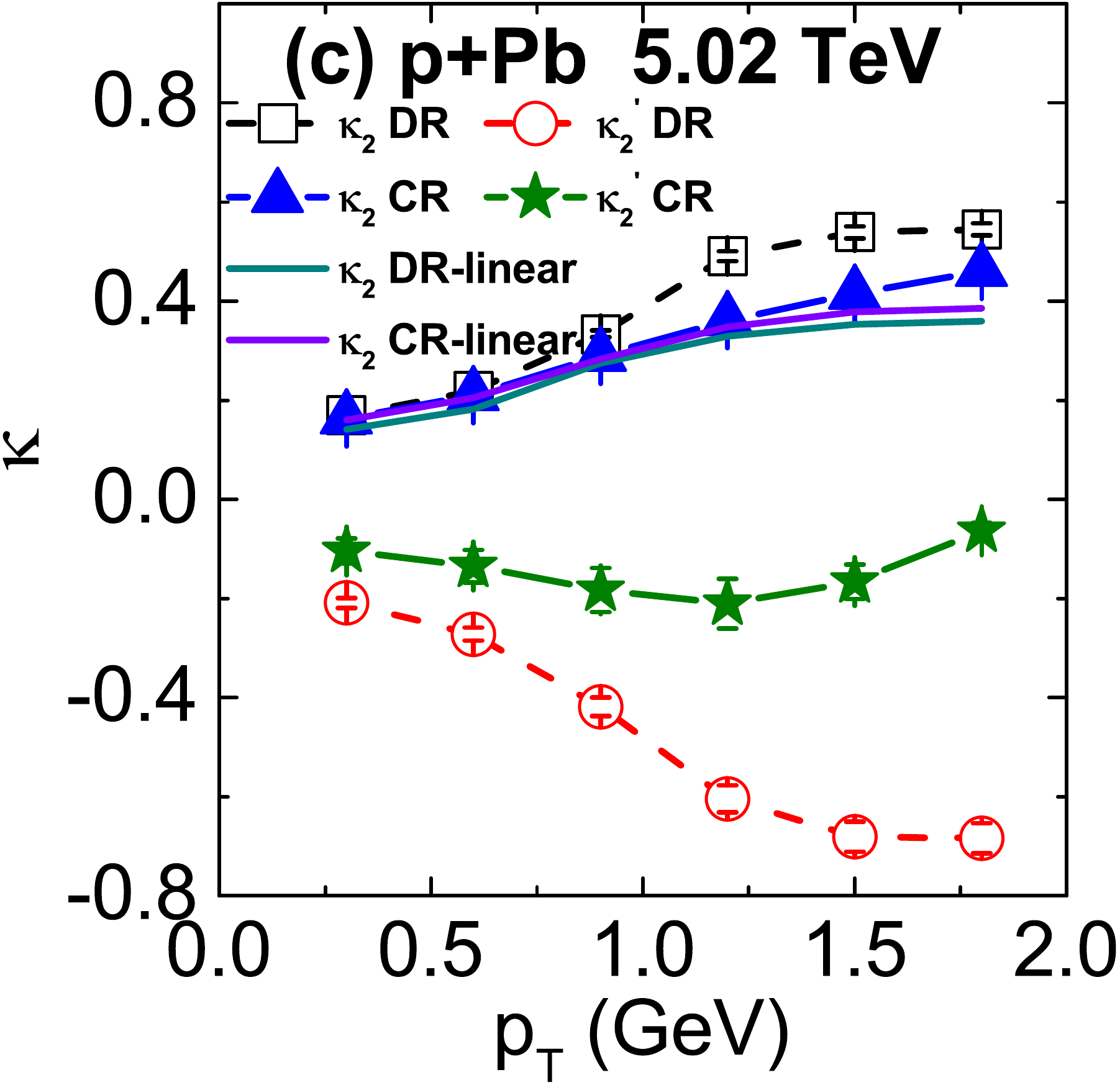}
\caption{(Color online)
Response relations $\kappa_2$ and
$\kappa_2'$ as a function of the transverse momentum, for Pb-Pb (30-40 \% and 70-80 \%) and p-Pb (b = 0-3 fm)
collisions at $\sqrt{s_{NN}}=5.02$ TeV, respectively. AMPT results are compared with ALICE data (based on the MC Glauber initial states) at 5.02 TeV~\cite{Acharya:2018afi} and hydrodynamic simulations [based on the MC-Kharzeev-Levin-Nardi(KLN) initial states] at 2.76 TeV~\cite{Qiu:2012hea}.
The results of linear response relations (dark cyan line and violet line) and linear+cubic response relations (symbol-line) are calculated by the DR and CR method in AMPT, respectively,
for (a) 30-40 \% Pb-Pb systems; (b) 70-80 \%Pb-Pb systems; (c) p-Pb (b=0-3 fm) systems.
}
\label{fig1}
\end{center}
\end{figure*}

\begin{figure}
\begin{center}
\includegraphics[width=0.35\textwidth]{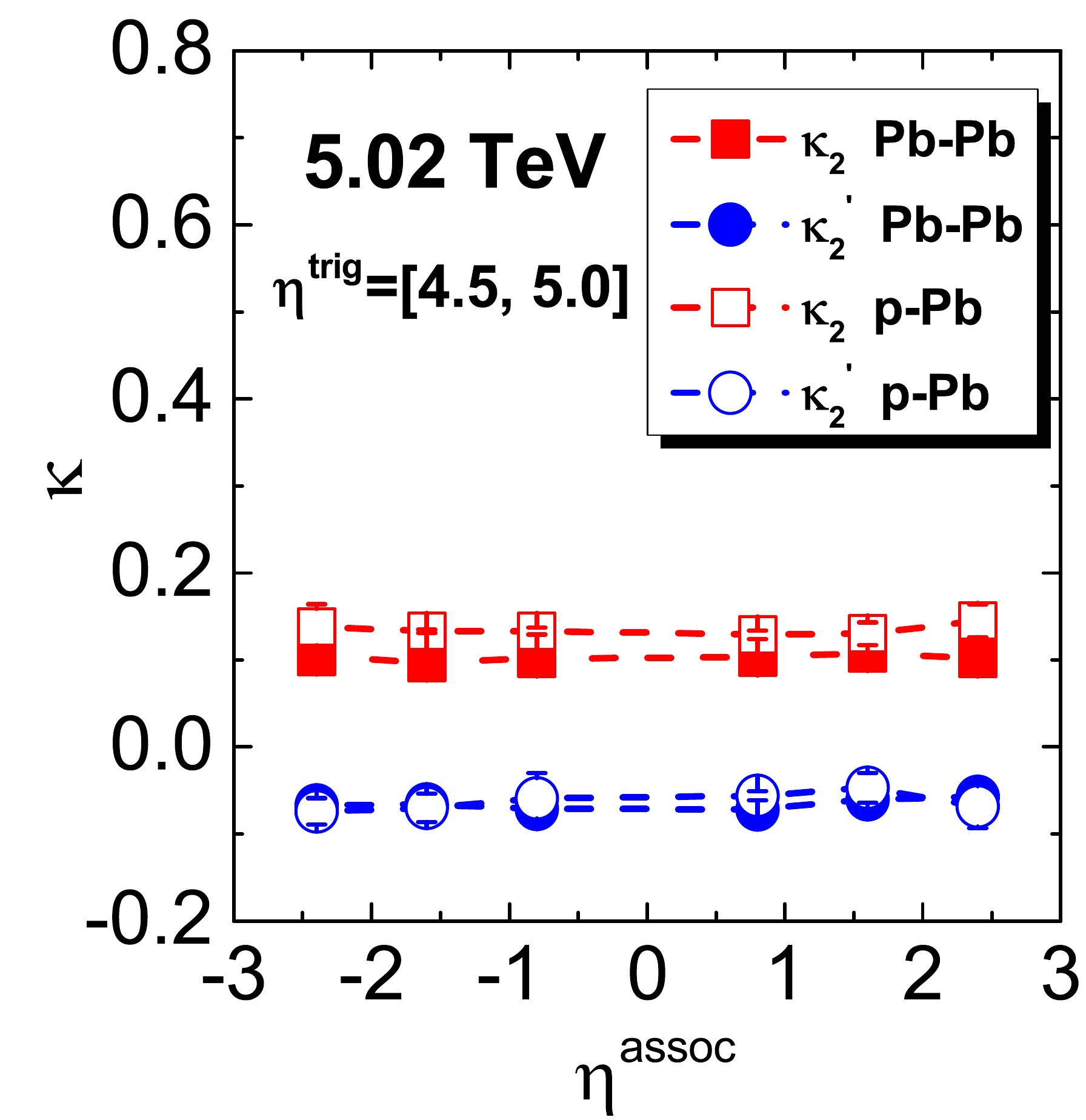}
\caption{(Color online)
Response relations $\kappa_2$ and
$\kappa_2'$ as a function of the pseudorapidity, for Pb-Pb (70-80 \%) and p-Pb (b = 0-3 fm)
collisions at $\sqrt{s_{NN}}=5.02$ TeV, respectively. The results of
$\kappa_2$ and $\kappa_2'$ (for linear+cubic response analyses) are calculated by the CR method.
}
\label{fig2}
\end{center}
\end{figure}

To compare the collective flow behavior in the two methods, $p_{T}$-related and $\eta$-related response relations
are introduced in \Fig{fig1} and \Fig{fig2}, respectively.
For simulations, we focus on Pb-Pb (30-40 \% and 70-80 \%) collisions and p-Pb (b=0-3 fm) collisions at $\sqrt{s_{NN}}=5.02$ TeV, respectively. \\

\Fig{fig1} shows that the response relations $\kappa_2$ and
$\kappa_2'$ as a function of the transverse momentum, from
AMPT simulations of Pb-Pb (30-40 \% and 70-80 \%) and p-Pb (b = 0-3 fm)
collisions at $\sqrt{s_{NN}}=5.02$ TeV, respectively.
The results of AMPT are compared with ALICE data~\cite{Acharya:2018afi} and hydrodynamic simulations~\cite{Qiu:2012hea} for 30-40 \% Pb-Pb collisions, are shown in \Fig{fig1} (a).
The ALICE data are results of $v_{2}\{2,|\Delta\eta|>2\}$ in 30-40 \% Pb-Pb collisions at 5.02 TeV~\cite{Acharya:2018afi} and a ratio with the initial eccentricity $\varepsilon_{2}\{2\}$, which is calculated by the Monte-Carlo (MC)-Glauber model. The hydrodynamic simulations at 2.76 TeV are based on the MC-(KLN) initial states.
The AMPT linear and linear+cubic response results agree with ALICE data for $v_{2}\{2,|\Delta\eta|>2\}/\varepsilon_{2}\{2\}$ (where $\varepsilon_{2}\{2\}$ is from the MC Glauber model)~\cite{Acharya:2018afi}, as shown in \Fig{fig1} (a).
One must note that only a linear response is considered in ALICE~\cite{Acharya:2018afi} and hydrodynamic simulations~\cite{Qiu:2012hea}, but linear response and linear+cubic response are both considered in the AMPT simulations.

For the linear response analyses, there is no significant difference in the linear response relations between the DR method and the CR method, shown in \Fig{fig1} (a), (b) and (c).
From~\Fig{fig1}, one can see that the linear $\kappa_2$ (dark cyan line and violet line) increases with transverse momentum increasing both in experimental data and theory simulations.
This effect should be the influence of radial flow as a consequence of the $p_{T}$ dependent response is controlled by the transverse density and size~\cite{Acharya:2018afi,Adam:2019ahi}.

For the linear+cubic response analyses, it makes sense that the cubic term is smaller than the linear term of the \Eq{eq:resp} simulations.
By the DR method, the cubic response is very weak in all the present systems. By the CR method, the response of the cubic item can be negligible in 30-40 \% Pb-Pb systems, but cannot be negligible in 70-80 \% Pb-Pb systems and p-Pb systems.
The three systems are compared in ~\Fig{fig1}, which shows the numerical solutions of the linear+cubic response relations, and the higher multiplicity systems (30-40 \% Pb-Pb collisions) and the lower multiplicity systems (70-80 \% Pb-Pb collisions and p-Pb (b=0-3~fm) collisions) are quite different.
These response relations calculated by the DR method and the CR method are almost identical in midcentral Pb-Pb collisions, but significantly different in lower multiplicity systems, i.e., the peripheral Pb-Pb collisions and p-Pb collisions.
Although non-hydrodynamic response has been subtracted in these response analyses,
multiplicity fluctuations may significantly influence the small medium response within the DR method and CR method.
The $\kappa_2$ in linear+cubic response increases with transverse momentum increases which is similar in linear response analyses.
The results of $\kappa_2'$ are quite different from the higher multiplicity systems [midcentral Pb-Pb collisions in \Fig{fig1} (a)]
and the lower multiplicity systems [peripheral Pb-Pb collisions in \Fig{fig1} (b) and p-Pb collisions in \Fig{fig1} (c)]. It shows a trend first increasing up to $p_T \sim$ 1.2 GeV, then
decreasing toward higher $p_T$, and less than zero when $p_T >$ 1.5 GeV in \Fig{fig1} (a).
Such negative $\kappa_2'$ at the present $p_{T}$ range on peripheral Pb-Pb collisions and p-Pb collisions are also shown in \Fig{fig1} (b) and \Fig{fig1} (c), respectively.
It has noted in the hydrodynamic model that such a simple cubic response as in \Eq{eq:resp} is not suited for the negative value of $\kappa_2'$ in low-multiplicity systems~\cite{Rao:2019bpo}.
The authors claimed that linear+cubic response no longer works well in small systems and one must consider other response types~\cite{Rao:2019bpo}.
However, it is still unclear what type of nonlinear response is compatible for such low-multiplicity systems.
These puzzles of cubic response in the lower multiplicity systems indicate that the current response analyses may need careful re-examination.
We will report our studies concerning these puzzles in the future.
Even though the initial fluctuation-driven cubic response $\kappa_2'$ is weak in the present Pb-Pb collisions (the value of integral $\kappa_2'$ is about 0.08 in 30-40 \% Pb-Pb collisions at 2.76 TeV which has been calculated by hydrodynamic~\cite{Noronha-Hostler:2015dbi} and AMPT models~\cite{De:2018hri}), it is still meaningful for studying the fluctuations of initial states.

Throughout \Fig{fig1}, the $\kappa_2$ in the linear response and in the linear+cubic response are almost identical where calculated by the DR and CR methods, except for in the lower multiplicity systems by the DR method.

Indeed, we also compare the $\kappa_2$ in the peripheral Pb-Pb collisions and p-Pb collisions. By the CR calculation, $p_{T}$-dependent response relations for the peripheral Pb-Pb systems and p-Pb  systems are similarly shown in \Fig{fig1} (b) and \Fig{fig1} (c),
which may imply that a collective response exists in the most central p-Pb collisions.  \\

The pseudorapidity related elliptic flow, $v_2(\eta)$,
provides information on the initial state and the early-time development of the collision,
constraining the description of the longitudinal dynamics in the collisions.
In practice, a pseudorapidity gap $|\Delta\eta| >$ 1 is
taken into account in our AMPT simulations for $v_2\{2, |\Delta\eta\}$.
For a single event, the particles pseudorapidity $\eta^{trig}$ ranges in [4.5, 5.0] are denoted as trigger particles, and one paired with pseudorapidity gap $|\Delta\eta| >$ 1 of the trigger particle is denoted associated particle (where within $|\eta^{assoc}| <$ 3.0 and each
has 0.8 unit of pseudorapidity bin size).
The linear+cubic CR response relations $\kappa_2$ and
$\kappa_2'$ as functions of the associated pseudorapidity, $\eta^{assoc}$, from AMPT simulations of Pb-Pb (70-80 \%) and p-Pb (b = 0-3 fm)
collisions at $\sqrt{s_{NN}}=5.02$ TeV are shown in \Fig{fig2}.
Here, we are focused on the $\eta$-independent linear+cubic response relations in the peripheral  Pb-Pb (70-80 \%) and the central p-Pb (b = 0-3 fm).
\Fig{fig2} shows a similar distribution of the response relations in peripheral Pb-Pb systems and p-Pb systems.
These response relations $\kappa_2$ and $\kappa_2'$ are smooth
at midpseudorapidity of the associated particles, while being different from the pseudorapidity-dependent response shown in Ref.~\cite{Hui:2019pdh}.
Note that the pseudorapidity-dependent response coefficients in Ref.~\cite{Hui:2019pdh} are local QGP fluctuations,
and our results act at the global QGP fluctuations, which are calculated by the cumulants method.
The behaviors persist at the large pseudorapidity gap, shown in \Fig{fig2},
which suggests that collective response exists in p-Pb collisions for the longitudinal expansions.
Similar $\eta$-independent results are found in the midcentral 30-40 \% Pb-Pb (not show in the figure), where $\kappa_2\approx 0.2$ and $\kappa_2' \approx 0.08$. \\

\begin{figure}
\begin{center}
\includegraphics[width=0.35\textwidth]{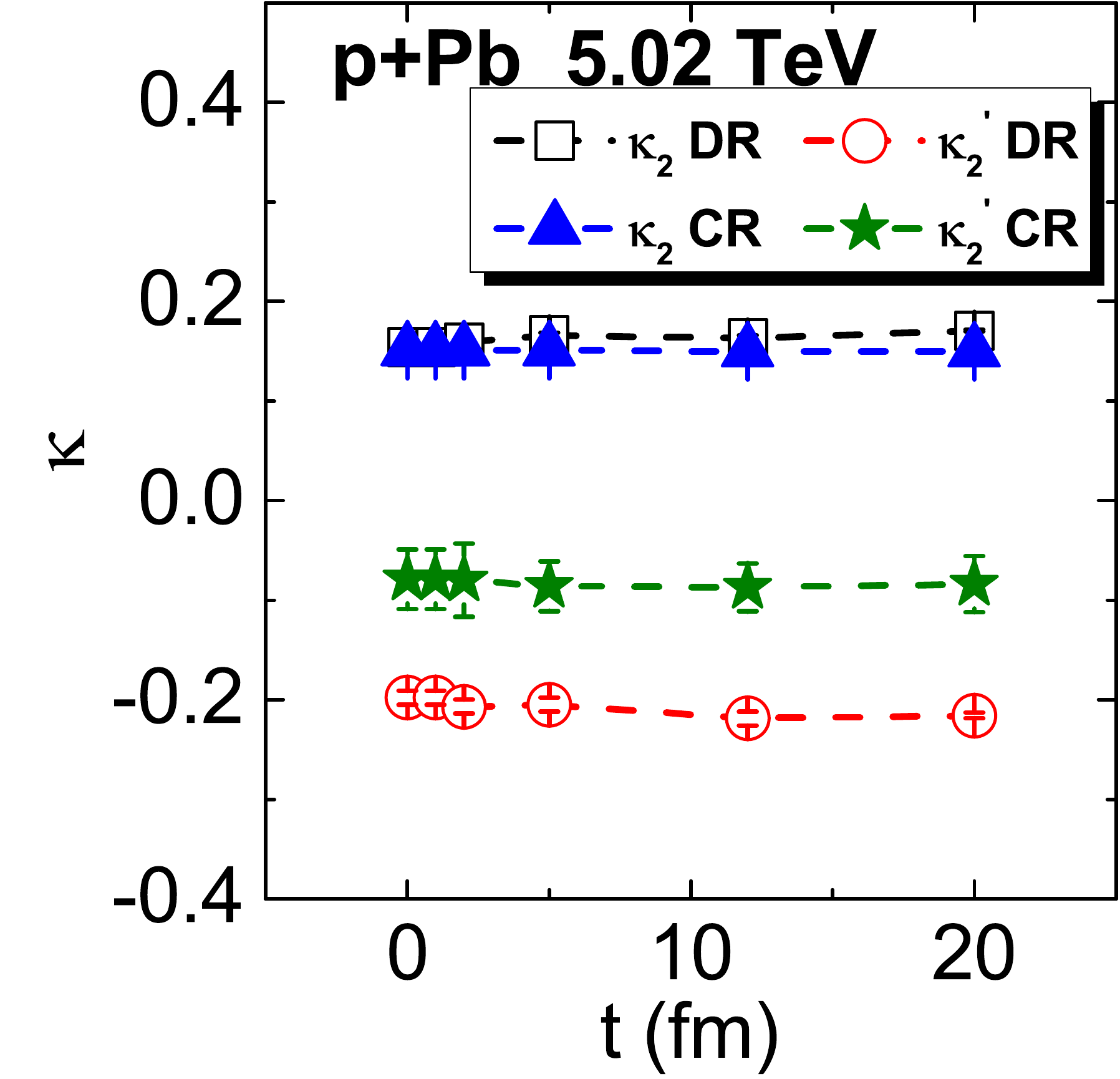}
\caption{(Color online)
Response relations $\kappa_2$ and
$\kappa_2'$ as a function of the hadron cascade time, for p-Pb
collisions at $\sqrt{s_{NN}}=5.02$ TeV. The results of
$\kappa_2$ and $\kappa_2'$ (for linear+cubic response analyses) are calculated by the DR and CR method, respectively.
}
\label{fig3}
\end{center}
\end{figure}

For Pb-Pb collisions, the collective flow mainly develops in the QGP phase~\cite{Song:2008cvh} and further accumulated in the hadronic stage~\cite{Chiho:2007seo}.
To determine the collective behavior in a small system, simulations on the response relations for different hadron cascade times are adopted.
Hadron cascade time is the maximum possible interaction time between a test hadron and other hadrons (excluding the interaction between partons and hadrons), that is, the time of the testing hadron from generation in hadron systems to kinetic freeze-out. These hadron cascade time can be obtained by setting the parameters in the subroutine ART of AMPT~\cite{Lin:2004en}.
\Fig{fig3} shows that $\kappa_2$ and $\kappa_2'$ as functions of the hadron cascade time, from AMPT simulations of p-Pb (b = 0-3 fm)
collisions at $\sqrt{s_{NN}}=5.02$ TeV, respectively.
The results of linear+cubic response relations are calculated by the DR and CR methods, respectively.
One can see that $\kappa_2$ is weakly depending on the hadron cascade time, and $\kappa_2'$ increases with the hadron cascade time increasing.
Again, the cubic response $\kappa_2'$ is dominated by initial fluctuations. Such effect of $\kappa_2'$ increases with the hadron cascade time increasing indicates that the final harmonic flow fluctuations can also be driven by the initial fluctuations.
The nonzero response relations calculated by the CR method in \Fig{fig3} implied that the contribution of collective response is dominated by the stage of medium expansions.
These results are consistent with a conclusion of Ref.~\cite{He:2016ape} for the AMPT model: that the collective mainly comes from the contribution of parton cascade and the hadron cascade weakly contributes. \\

\section{Summary}
\label{sec:sum}

In summary, we have carried out event-by-event AMPT simulations for Pb-Pb (30-40 \% and 70-80 \%) collisions and p-Pb (b=0-3~fm) collisions at $\sqrt{s_{NN}}=5.02$ TeV, respectively.
Base on calculation of the directed response (DR) method and the cumulants response (CR) method, the response relations as a function of the transverse momentum, are both shown in Pb-Pb and p-Pb collisions. Such $p_{T}$-dependent $\kappa_2$ in linear response and linear+cubic response agree with ALICE data for $v_{2}\{2,|\Delta\eta|>2\}/\varepsilon_{2}\{2\}$ (where $\varepsilon_{2}\{2\}$ is based on MC Glauber).
By comparing the DR and CR methods, we found that the linear response relations are almost identical in all the present collisions. Similar results of linear+cubic response relations
are also shown in the higher multiplicity systems (30-40 \% Pb-Pb collisions). It has become a significant difference in the lower multiplicity systems, i.e., the peripheral Pb-Pb collisions and p-Pb collisions.
Throughout the whole $p_{T}$-dependent simulations, the $\kappa_2$ in the linear response and in the linear+cubic response are almost identical where calculated by the DR and CR methods, except for in the lower multiplicity systems calculated by the DR method. The cubic response is much weaker than the linear response in the simulations, except for in the lower multiplicity systems calculated by the DR method.
The linear+cubic response may not suit for the lower multiplicity systems by the hydrodynamic model,
but it is still unclear what types of nonlinear response are compatible for such small multiplicity systems.
Furthermore, $p_{T}$-dependent and $\eta$-independent response relations with the CR calculation are similarly shown in peripheral Pb-Pb systems and p-Pb systems,
which may imply that a collective response exists in the most central p-Pb collisions.
These collective behaviors are predominantly produced when the medium expands.

\section*{Acknowledgements}
We thank L. Yan, H. Li and J. Noronha-Hostler for very helpful suggestions and discussions. D.-X. W. has supported by the Youth Program of Natural Science Foundation of Guangxi (China) Grant No.~2019GXNSFBA245080, the Special fund for talents of Guangxi (China) Grant No.~AD19245157, and the Doctor Startup Foundation of Guangxi University of Science and Technology Grant No.~19Z19. L.-J. Z. has supported by the National Natural Science Foundation of China Grant No.~11865005 and the Natural Science Foundation of Guangxi (China) Grant No.~2018GXNSFAA281024.


\begin{thebibliography}{99}
\bibitem{Abelev:2009lrr}
      B.~Abelev, M.~M.~Aggarwal, Z.~Ahammed, A.~V.~Alakhverdyants, B.~D.~Anderson, D.~Arkhipkin, G.~S.~Averichev, J.~Balewski, O.~Barannikova, L.~S.~Barnby, {\it et al.} [STAR Collaboration],
      Phys.\ Rev.\ C {\bf 80}, 064912 (2009)
      [arXiv:0909.0191 [nucl-ex]].

\bibitem{ATLAS:2012mot}
      G.~Aad, B.~Abbott, J.~Abdallah, S.~Abdel Khalek, A.~A.~Abdelalim, A.~Abdesselam, O.~Abdinov, B.~Abi, M.~Abolins, O.~S.~AbouZeid, {\it et al.} [ATLAS Collaboration],
      Phys.\ Rev.\ C {\bf 86}, 014907 (2012)
      [arXiv:1203.3087 [hep-ex]].
\bibitem{McDonald:2017hpf}
      S.~McDonald, C.~Shen, F. Fillion-Gourdeau, S. Jeon, C. Gale,
      Phys.\ Rev.\ C {\bf 95}, 064913 (2017)
      [arXiv:1609.02958 [hep-ph]].

\bibitem{Alver:2010tfi}
      B.~Alver, C.~Gombeaud, M.~Luzum, and J.-Y.~Ollitrault,
      Phys.\ Rev.\ C {\bf 82}, 034913 (2010)
      [arXiv:1609.02958 [hep-ph]].

\bibitem{Noronha-Hostler:2015dbi}
      J.~Noronha-Hostler, L.~Yan, F.~G.~Gardim and J.-Y.~Ollitrault,
      Phys.\ Rev.\ C {\bf 93}, 014909 (2016)
      [arXiv:1511.03896 [nucl-th]].
\bibitem{Yan:2017ivm}
      L.~Yan,
      Chin.\ Phys.\ C {\bf 42}, 042001 (2018)
      [arXiv:1712.04580 [nucl-th]].
\bibitem{De:2018hri}
      D.-X.~Wei, X.-G.~Huang and L.~Yan,
      Phys.\ Rev.\ C {\bf 98}, 044908 (2018)
      [arXiv:1807.06299 [nucl-th]].

\bibitem{Ma:2016ipc}
      L.~Ma, G.~L.~Ma, Y.~G.~Ma,
      Phys.\ Rev.\ C {\bf 94}, 044915 (2016)
      [arXiv:1610.04733 [nucl-th]].

\bibitem{Hui:2019pdh}
      H.~Li and L. Yan,
      Phys.\ Lett.\ B {\bf 802}, 135248 (2020)
      [arXiv:1907.10854 [nucl-th]].

\bibitem{CMS:2017efc}
      V.~Khachatryan, A.M.~Sirunyan, A.~Tumasyan, W.~Adam, E.~Asilar, T.~Bergauer, J.~Brandstetter, E.~Brondolin, M.~Dragicevic, J.~Er$\ddot{o}$, {\it et al.} [CMS Collaboration],
      Phys.\ Lett.\ B {\bf 765}, 193 (2017)
      [arXiv:1606.06198 [nucl-ex]].

\bibitem{CMS:2015efc}
      V.~Khachatryan, A.M.~Sirunyan, A.~Tumasyan, W.~Adam, T.~Bergauer, M.~Dragicevic, J.~Er$\ddot{o}$, M.~Friedl, R.~Fr$\ddot{u}$hwirth, V.M.~Ghete, {\it et al.} [CMS Collaboration],
      Phys.\ Rev.\ Lett.\ {\bf 115}, 012301 (2015)
      [arXiv:1502.05382 [nucl-ex]].

\bibitem{ATLAS:2018mol}
      M.~Aaboud, G.~Aad, B.~Abbott, O.~Abdinov, B.~Abeloos, S.~H.~Abidi, O.~S.~AbouZeid, N.~L.~Abraham, H.~Abramowicz, H.~Abreu, {\it et al.} [ATLAS Collaboration],
      Phys.\ Rev.\ C {\bf 97}, 024904 (2018)
      [arXiv:1708.03559 [hep-ex]].
\bibitem{PHENIX:2018mom}
      C.~Aidala, Y.~Akiba, M.~Alfred, V.~Andrieux, K.~Aoki, N.~Apadula, H.~Asano, C.~Ayuso, B.~Azmoun, V.~Babintsev, {\it et al.} [PHENIX Collaboration],
      Phys.\ Rev.\ Lett.\ {\bf 120}, 062302 (2018)
      [arXiv:1707.06108 [nucl-ex]].
\bibitem{Nagle:2018ssc}
      J.~L.~Nagle and W.~A.~Zajc,
      Ann.\ Rev.\ Nucl.\ Part.\ Sci.\ {\bf 68}, 0211 (2018)
      [arXiv:1801.03477 [nucl-ex]].
\bibitem{Dusling:2016ncp}
      K.~Dusling, W.~Li and B.~Schenke,
      Int.\ J.\ Mod.\ Phys.\ E {\bf 25}, 1630002 (2016)
      [arXiv:1509.07939 [nucl-ex]].
\bibitem{Schenke:2017ooc}
      B.~Schenke,
      Nucl.\ Phys.\ A {\bf 967}, 105 (2017)
      [arXiv:1704.03914 [nucl-th]].
\bibitem{Lin:2004en}
      Z.-W.~Lin, C.~Ko, B.-A.~Li, B.~Zhang and S.~Pal,
      Phys.\ Rev.\ C {\bf 72}, 064901 (2005)
      [arXiv:nucl-th/0411110].
\bibitem{Alver:2010gr}
      B.~Alver and G.~Roland,
      Phys.\ Rev.\ C {\bf 81}, 054905 (2010)
      [Erratum: Phys. Rev.C82,039903(2010)], [arXiv:1003.0194 [nucl-th]].
\bibitem{Heinz:2013cfa}
      U.~Heinz and R.~Snellings,
      Ann.\ Rev.\ Nucl.\ Part.\ Sci.\ {\bf 63}, 123 (2013)
      [arXiv:1301.2826 [nucl-th]].
\bibitem{ALICE:2016cef}
      J.~Adam, D.~Adamov$\acute{a}$, M.M.~Aggarwal, G.~Aglieri~Rinella, M.~Agnello, N.~Agrawal, Z.~hammed, S.~Ahmad, S.U.~Ahn, S.~Aiola, {\it et al.} [ALICE Collaboration],
      Phys.\ Rev.\ Lett.\ {\bf 117}, 182301 (2016)
      [arXiv:1604.07663 [nucl-ex]].
\bibitem{Nie:2019ioi}
      M.~Nie, L.~Yi, X.~Luo, G.~Ma and J.~Jia ,
      Phys.\ Rev.\ C {\bf 100}, 064905 (2019)
      [arXiv:1906.01422 [nucl-th]].
\bibitem{Teaney:2010vd}
      D.~Teaney and L.~Yan,
      Phys.\ Rev.\ C {\bf 83}, 064904 (2011)
      [arXiv:1010.1876 [nucl-th]].

\bibitem{Chatrchyan:2014kba}
      S.~Chatrchyan, V.~Khachatryan, A.M.~Sirunyan, A.~Tumasyan, W.~Adam, T.~Bergauer, M.~Dragicevic, J.~Er$\ddot{o}$, C.~Fabjan, M.~Friedl, {\it et al.} [CMS Collaboration],
      Phys.\ Rev.\ C {\bf 89}, 044906 (2014)
      [arXiv:1310.8651 [nucl-ex]].
\bibitem{Acharya:2018zuq}
      S.~Acharya, F.T.-.~Acosta, D.~Adamov$\acute{a}$, J.~Adolfsson, M.M.~Aggarwal, G.~Aglieri Rinella, M.~Agnello, N.~Agrawal, Z.~Ahammed, S.U.~Ahn, {\it et al.} [ALICE Collaboration],
      JHEP {\bf 09}, 006 (2018)
      [arXiv:1805.04390 [nucl-ex]].
\bibitem{Aad:2014vba}
      G.~Aad, B.~Abbott, J.~Abdallah, S.~Abdel Khalek, O.~Abdinov, R.~Aben, B.~Abi, M.~Abolins, O.S.~AbouZeid, H.~Abramowicz, {\it et al.} [ATLAS Collaboration],
      Eur.\ Phys.\ J.\ C {\bf 74}, 3157 (2014)
      [arXiv:1408.4342 [hep-ex]].
\bibitem{Chatrchyan:2013mot}
      S.~Chatrchyan, V.~Khachatryan, A.M.~Sirunyan, A.~Tumasyan, W.~Adam, T.~Bergauer, M.~Dragicevic, J.~Er$\ddot{o}$, C.~Fabjan, M.~Friedl, {\it et al.} [CMS Collaboration],
      Phys.\ Rev.\ C {\bf 87}, 014902 (2013)
      [arXiv:1204.1409 [nucl-ex]].
\bibitem{Yan:2014aad}
      L.~Yan, J.-Y.~Ollitrault and A.~Poskanzer,
      Phys.\ Lett.\ B {\bf 742}, 290 (2015)
      [arXiv:1408.0921 [nucl-th]].

\bibitem{Acharya:2018afi}
      S.~Acharya, F.T.-.~Acosta, D.~Adamov$\acute{a}$, J.~Adolfsson, M.M.~Aggarwal, G.~Aglieri Rinella, M.~Agnello, N.~Agrawal, Z.~Ahammed, S.U.~Ahn, {\it et al.} [ALICE Collaboration],
      Phys.\ Lett.\ B {\bf 784}, 82 (2018)
      [arXiv:1805.01832 [nucl-ex]].
\bibitem{Qiu:2012hea}
      Zhi Qiu, Chun Shen, Ulrich W. Heinz,
      Phys.\ Lett.\ B {\bf 707}, 151 (2012)
      [arXiv:1110.3033 [nucl-th]].

\bibitem{Adam:2019ahi}
      J.~Adam, L.~Adamczyk, J.~R.~Adams, J.~K.~Adkins, G.~Agakishiev, M.~M.~Aggarwal, Z.~Ahammed, I.~Alekseev, D.~M.~Anderson, R.~Aoyama, {\it et al.} [STAR Collaboration],
      Phys.\ Rev.\ Lett.\ {\bf 122}, 172301 (2019)
      [arXiv:1901.08155 [nucl-ex]].

\bibitem{Rao:2019bpo}
      S.~Rao, M. Sievert and J.~Noronha-Hostler,
      [arXiv:1910.03677 [nucl-th]].


\bibitem{Song:2008cvh}
      H.~Song U.~W.~Heinz,
      Phys.\ Rev.\ C {\bf 77}, 064901 (2008)
      [arXiv:0712.3715 [nucl-th]].
\bibitem{Chiho:2007seo}
      C.~Nonaka and S.~A.~Bass,
      Phys.\ Rev.\ C {\bf 75}, 014902 (2007)
      [arXiv:0607018 [nucl-th]].

\bibitem{He:2016ape}
      L.~He, T.~Edmonds, Z.-W~Lin, F.~Liu, D.~Molnar, and F.~Wang
      Phys.\ Lett.\ B {\bf 753}, 506 (2016)
      [arXiv:1502.05572 [nucl-th]].



\end{thebibliography}

\end{document}